\documentclass[twocolumn,english,prl]{revtex4}
\usepackage[T1]{fontenc}
\usepackage[latin9]{inputenc}
\usepackage{color}
\usepackage{babel}
\usepackage{amsmath}
\usepackage{amssymb}
\usepackage{graphicx}
\usepackage[unicode=true,pdfusetitle,
 bookmarks=true,bookmarksnumbered=false,bookmarksopen=false,
 breaklinks=false,pdfborder={0 0 1},backref=section,colorlinks=true]
 {hyperref}

\makeatletter
\@ifundefined{textcolor}{}
{%
 \definecolor{BLACK}{gray}{0}
 \definecolor{WHITE}{gray}{1}
 \definecolor{RED}{rgb}{1,0,0}
 \definecolor{GREEN}{rgb}{0,1,0}
 \definecolor{BLUE}{rgb}{0,0,1}
 \definecolor{CYAN}{cmyk}{1,0,0,0}
 \definecolor{MAGENTA}{cmyk}{0,1,0,0}
 \definecolor{YELLOW}{cmyk}{0,0,1,0}
}

\usepackage{babel}

\usepackage{amsfonts}\usepackage{bbm}\usepackage{graphics}

\def\ket#1{\left| #1\right>}
\def\bra#1{\left< #1\right|}

\def\<{\langle}
\def\>{\rangle}

\usepackage{leftidx}

\usepackage{babel}

\makeatother

\begin{document}

\title{Tunable Spin Qubit Coupling Mediated by a Multi-Electron Quantum
Dot}

\author{V. Srinivasa}

\email{vsriniv@umd.edu}

\affiliation{Joint Quantum Institute, University of Maryland, College Park, MD
20742, USA and National Institute of Standards and Technology, Gaithersburg,
MD 20899, USA}

\author{H. Xu}

\affiliation{Joint Quantum Institute, University of Maryland, College Park, MD
20742, USA and National Institute of Standards and Technology, Gaithersburg,
MD 20899, USA}

\author{J. M. Taylor}

\affiliation{Joint Quantum Institute, University of Maryland, College Park, MD
20742, USA and National Institute of Standards and Technology, Gaithersburg,
MD 20899, USA}
\begin{abstract}
We present an approach for entangling electron spin qubits localized
on spatially separated impurity atoms or quantum dots via a multi-electron,
two-level quantum dot. The effective exchange interaction mediated
by the dot can be understood as the simplest manifestation of Ruderman-Kittel-Kasuya-Yosida
exchange, and can be manipulated through gate voltage control of level
splittings and tunneling amplitudes within the system. This provides
both a high degree of tuneability and a means for realizing high-fidelity
two-qubit gates between spatially separated spins, yielding an experimentally
accessible method of coupling donor electron spins in silicon via
a hybrid impurity-dot system. 
\end{abstract}
\maketitle
Single spins in solid-state systems represent versatile candidates
for scalable quantum bits (qubits) in quantum information processing
architectures \cite{Loss1998,Kane1998,Hanson2007RMP,Hanson2008,Awschalom2013,Zwanenburg2013}.
In many proposals involving single-spin qubits localized on impurity
atoms \cite{Kane1998,Vrijen2000} and within quantum dots \cite{Loss1998,Burkard1999},
two-qubit coupling schemes harness the advantages of tunneling-based
nearest-neighbor exchange interactions: exchange gates are rapid,
tunable, and protected against multiple types of noise \cite{DiVincenzo2000Nature,Wu2002,Taylor2005,Doherty2013,Taylor2013}.
These features have been demonstrated for electron spins in quantum
dots \cite{Petta2005,Maune2012,Medford2013NNano,Medford2013}, while
a similar demonstration for spins localized on impurity atoms such
as phosphorus donors in silicon remains an outstanding experimental
challenge \cite{Zwanenburg2013}. Although the exchange interaction
originates from the long-range Coulomb interaction, its strength typically
decays exponentially with distance \cite{Herring1964,Burkard1999}.
Long-range coupling via concatenation of multiple nearest-neighbor
interactions is not ideal for coupling spatially separated electron
spins, as it sets a low threshold error rate below which fault-tolerant
quantum computing is feasible \cite{Szkopek2006,Stephens2009}. A
mechanism for long-range coupling that simultaneously enables scalability
and robustness against errors is therefore key to realizing practical
spin-based quantum information processing devices.

Approaches to implementing long-range interactions typically involve
identifying a system that acts as a mediator of the interaction between
the qubits, with proposed systems including optical cavities and microwave
stripline resonators \cite{Imamoglu1999,Childress2004,Burkard2006,Taylor2006,Frey2012,Petersson2012},
floating metallic \cite{Trifunovic2012} and ferromagnetic \cite{Trifunovic2013}
couplers, the collective modes of spin chains \cite{Friesen2007,Srinivasa2007,Oh2010},
superconducting systems \cite{Marcos2010,Leijnse2013}, and multi-electron
molecular cores \cite{Lehmann2007}. Recently, long-range coupling
of electrons located in the two outer quantum dots of a linear triple
dot system has been demonstrated \cite{Busl2013,Braakman2013}. The
effective exchange interaction in that system arises from electron
cotunneling between the outer dots and exhibits the fourth-order dependence
on tunneling amplitudes that is characteristic of superexchange \cite{Recher2001},
but suffers from a large virtual energy cost from the doubly occupied
center dot states. In contrast, a many-electron quantum dot in the
center can also couple distant spins via the Ruderman-Kittel-Kasuya-Yosida
(RKKY) interaction, with low-energy intermediate states \cite{Craig2004},
but perhaps at the cost of low fidelity as impurity-Fermi sea correlations
become hard to disentangle.

Here, we show that a multi-level quantum dot containing two electrons
can mediate a high-fidelity exchange interaction between two spatially
separated single-electron spin qubits. We assume in what follows that
the qubit electrons are localized on single-level impurity atoms,
but our analysis also maps directly to the case of a triple quantum
dot system with the same level structure and electron occupation.
Our approach suggests an experimentally accessible method for achieving
tunable coupling between donor electron spins in silicon \cite{Zwanenburg2013,Morello2010,Pla2012}. 

\begin{figure}
\includegraphics[width=3.375in]{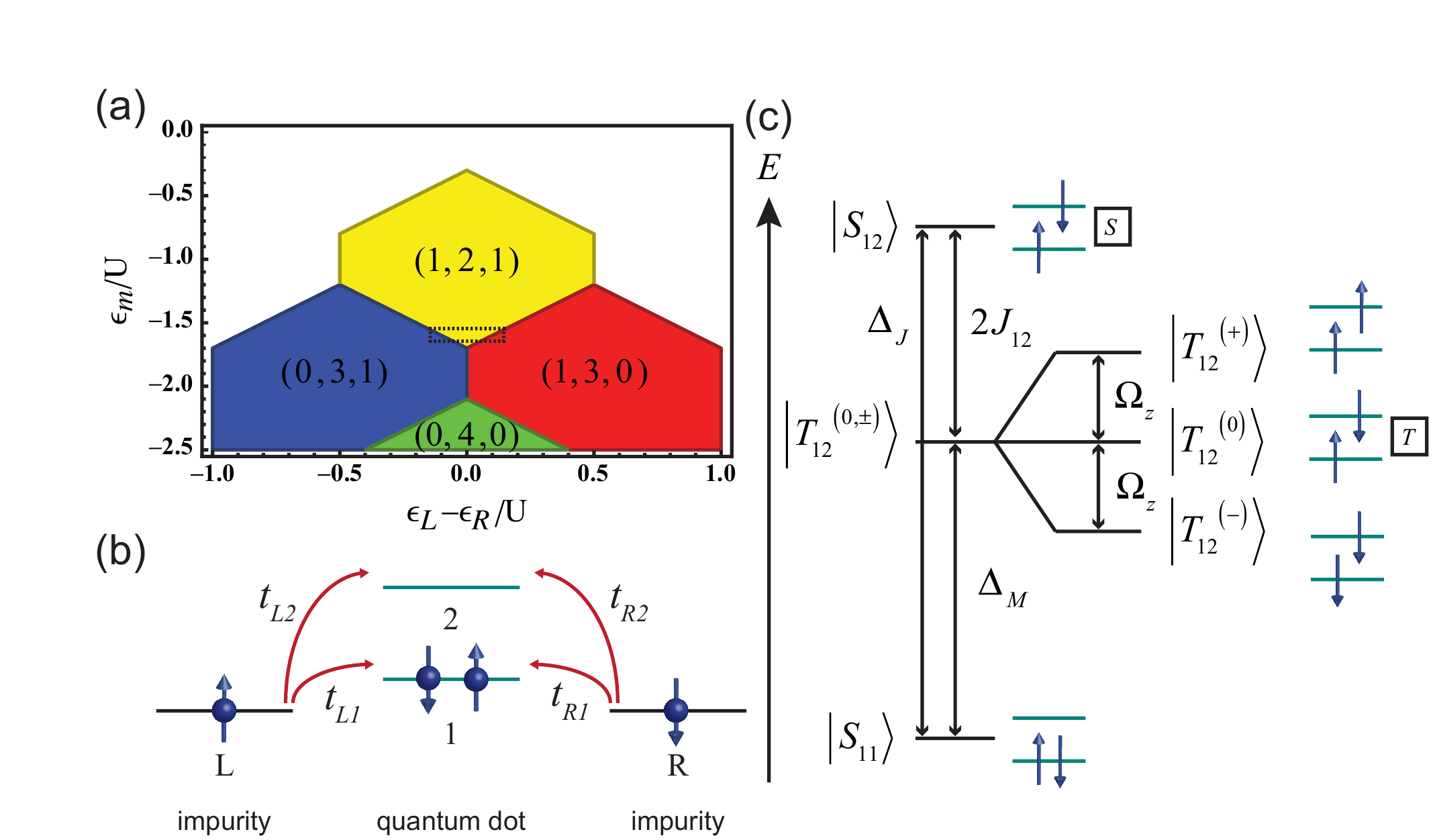}

\caption{\label{fig:leveldiag} (a) Charge stability diagram for the combined
impurity-dot three-site model ($U\equiv U_{1},$ $\epsilon_{m}\equiv\epsilon_{1}$),
with the operating point indicated. (b) Schematic diagram showing
the orbitals of a pair of single-level impurity atoms coupled via
a two-level quantum dot. The electron occupation illustrates the initial
configuration $\left(1,2,1\right)$. Arrows depict the tunneling amplitudes
defined in Eq. (\ref{eq:Ht}). Reversing the direction of an arrow
corresponds to taking the complex conjugate of the associated tunneling
amplitude. (c) Energy level diagram illustrating the two-spin states
of the mediator dot used in our calculation. }
\end{figure}

\emph{Hubbard model description}: The minimal model for our approach
comprises a two-level quantum dot coupled to two impurities which
are chosen to be near their ionization point by appropriate choice
of gate voltages. This reduces to a multi-orbital Hubbard model for
a linear three-site system in the four-electron regime \cite{Korkusinski2007,Hsieh2012}.
We assume gate voltages can be applied to the system such that the
total electron number can be set to be four, while the charge stability
diagram prefers the initial configuration of $\left(1,2,1\right)$.
Here, $\left(n_{L},n_{M},n_{R}\right)$ represents the configuration
with $n_{L}$ ($n_{R}$) electrons in impurity orbital $L$ ($R$)
and $n_{M}$ electrons in the mediator dot (Fig. \ref{fig:leveldiag}).
We work at a point in the charge stability diagram where transitions
to charge configurations $\left(0,3,1\right)$ and $\left(1,3,0\right)$
are the closest available charge states, with detunings $\Delta_{L},\Delta_{R}$
{[}Fig. \ref{fig:leveldiag}(a){]}.

We can write the Hamiltonian as $H_{{\rm hub}}=H_{n}+H_{t},$ where
\begin{eqnarray}
H_{n} & = & \sum_{i}\epsilon_{i}n_{i}+\frac{U_{i}}{2}n_{i}(n_{i}-1)+\sum_{i\neq j}\frac{K_{ij}}{2}n_{i}n_{j}\nonumber \\
 &  & +J_{12}\sum_{\sigma,\sigma'}c_{1,\sigma}^{\dag}c_{2,\sigma'}^{\dag}c_{1,\sigma'}c_{2,\sigma},\label{eq:Hn}\\
H_{t} & = & -\sum_{i=1,2}\sum_{\sigma}\left(t_{Li}c_{i,\sigma}^{\dag}c_{L,\sigma}+t_{Ri}c_{i,\sigma}^{\dag}c_{R,\sigma}+{\rm h.c.}\right)\label{eq:Ht}
\end{eqnarray}
with $i,j=L,R,1,2$ denoting the left ($L$) and right ($R$) impurity
orbital levels and the lower-energy (1) and higher-energy (2) orbitals
of the center quantum dot {[}Fig. \ref{fig:leveldiag}(b){]}. $H_{n}$
is diagonal with respect to the charge occupation defined by the set
of eigenvalues of the electron number operators $n_{i}=\sum_{\sigma}n_{i,\sigma}=\sum_{\sigma}c_{i,\sigma}^{\dagger}c_{i,\sigma},$
where $c_{i,\sigma}^{\dagger}$ creates an electron in orbital $i$
with spin $\sigma.$ The quantity $\epsilon_{i}$ denotes the on-site
energy of orbital $i$. $U_{i}$ and $K_{ij}$ are the Coulomb repulsion
energies for two electrons in the same orbital $i$ and in different
orbitals $i$ and $j,$ respectively, and $J_{12}$ is the exchange
energy for electrons in orbitals 1 and 2 of the dot with spins $\sigma,\sigma'=\uparrow,\downarrow.$

Since we assume at most single occupancy of the impurity levels $L$
and $R$ and a linear geometry for the three sites, we implicitly
have set $U_{L},U_{R}\rightarrow\infty$ and have neglected $K_{LR}$
in Eq. (\ref{eq:Hn}). We also neglect the term involving $U_{2},$
as configurations such as $\left(0,4,0\right)$ which involve double
occupancy of orbital 2 are both high in energy and not well justified
within a two-orbital picture for the dot in the presence of the Coulomb
interaction. Additionally, we assume symmetric Coulomb repulsion energies
between the impurities and the dot and set $K_{Li}=K_{Ri}\equiv K_{i}$
for $i=1,2,$ while we take exchange terms $J_{Ri}=J_{Li}=0$, appropriate
for weak tunneling. The tunneling term $H_{t}$ couples subspaces
of fixed charge occupation and is expressed in terms of the complex
tunneling amplitudes $t_{Li,Ri}$ between orbitals $L,R$ and orbital
$i$ of the dot. Note that we define $t_{Li,Ri}$ as the amplitudes
for tunneling from the outer sites into the center dot and $t_{Li,Ri}^{\ast}$
as the amplitudes for tunneling in the opposite direction {[}see Fig.
\ref{fig:leveldiag}(b){]}.

\begin{figure}
\includegraphics[bb=0bp 20bp 526bp 372bp,width=3.375in]{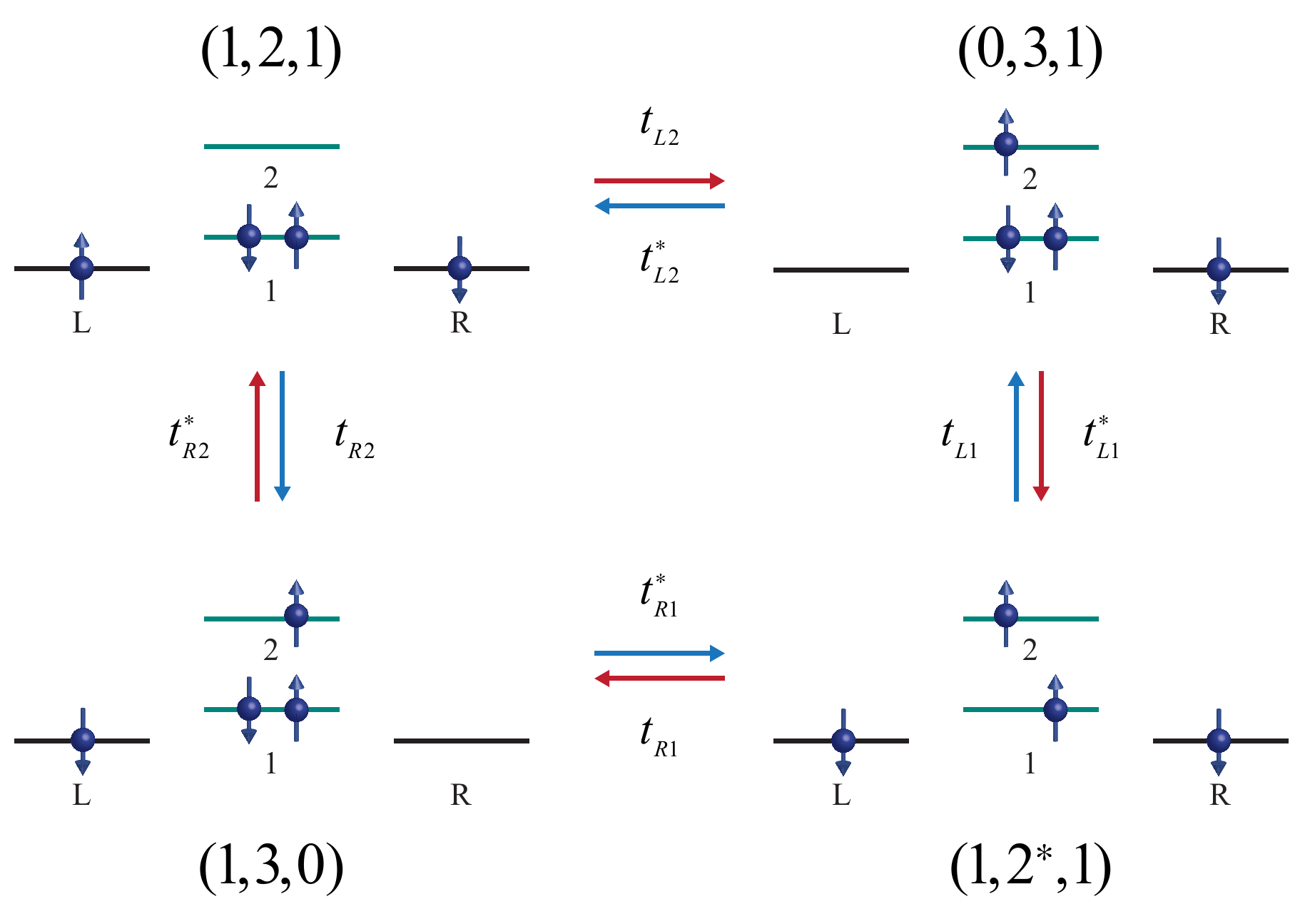}

\caption{\label{fig:vtp}Schematic illustration of virtual tunneling processes
which give rise to the effective exchange interaction in Eq. \eqref{eq:J}.
The red (blue) arrows correspond to the process in which the electron
in orbital $L$ ($R$) tunnels to the center dot in the first step.
Each step is labeled with the tunnel coupling for the associated hopping
term in $H_{t}$ {[}Eq. (\ref{eq:Ht}){]}. }
\end{figure}

In the present work, we are interested in a system where we can effectively
turn on and off the induced exchange, either by gate voltage (varying
the energy difference between different charge sectors) or by tuning
tunneling. We consider our low-energy manifold to be the $\left(1,2,1\right)$
charge configuration with the center dot spins in the lowest-energy
singlet. This set of states is gapped (as shown below) from other
configurations by an energy large compared to typical dilution refrigerator
temperatures and provides the starting point for our perturbation
theory, in which we take $H_{t}$ as a perturbation to $H_{n}.$ To
further simplify the calculation, we note that the Hubbard Hamiltonian
$H_{{\rm hub}}$ conserves both the total spin $S_{\textrm{tot}}$
and the total $z$ component of spin $S_{z}$ for the four-electron
system. Thus, we can independently consider the two subspaces $\left(S_{\textrm{tot}}=0,S_{z}=0\right)$
and $\left(S_{\textrm{tot}}=1,S_{z}=0\right).$ Since $H_{t}$ is
independent of spin, the set of charge configurations generated by
applying $H_{t}$ to $\left(1,2,1\right)$ is identical for these
two spin subspaces. Neglecting configurations which involve double
occupancy of orbitals $L,$ $R,$ and 2, the intermediate charge configurations
generated by $H_{t}$ are $\left(0,3,1\right),$ $\left(1,3,0\right),$
and $\left(1,2^{\ast},1\right),$ where $n_{M}=2^{*}$ denotes an
excited two-electron state of the dot in which one electron is in
orbital 1 and the second electron is in orbital 2 (see Fig. \ref{fig:vtp}).
According to Eq. (\ref{eq:Hn}), the energies of the $\left(1,2,1\right)$
states are identical in the absence of tunneling via the center dot
and are equal to $E_{0}=\epsilon_{L}+\epsilon_{R}+2\epsilon_{1}+U_{1}+4K_{1}.$
Choosing $E_{0}$ as the energy origin, we can determine the zeroth-order
energies of the intermediate states from Eq. (\ref{eq:Hn}). The energies
of the $\left(0,3,1\right)$ {[}$\left(1,3,0\right)${]} states are
\begin{equation}
\Delta_{L\left(R\right)}=\epsilon_{2}-\epsilon_{L\left(R\right)}+W\ ,
\end{equation}
where $W\equiv-2K_{1}+K_{2}+2K_{12}-J_{12}.$ The energy of each $\left(1,2^{\ast},1\right)$
state has one of two values, depending on the two-spin state of the
center dot electrons: for the triplet and singlet states, the energies
{[}Fig. \ref{fig:leveldiag}(c){]} are, respectively, 
\begin{align}
\Delta_{M} & =\epsilon_{2}-\epsilon_{1}+W-U_{1}+K_{2}-K_{12}\ ,\\
\Delta_{J} & =\Delta_{M}+2J_{12}\ .
\end{align}

Within our toy model, the effective exchange coupling is given by
the energy splitting between the states $\ket{\left(1,2,1\right);S_{LR},S_{11}}$
and $\ket{\left(1,2,1\right);T_{LR}^{\left(0\right)},S_{11}}$ in
the presence of the tunneling term $H_{t}.$ Here, $\ket{S_{ij}}$
and $\ket{T_{ij}^{\left(m\right)}}$ represent two-electron singlet
and triplet spin states of the electrons in orbitals $i,j$ and $m=0,\pm$
indicates the spin magnetic quantum number of the triplet state. Since
there is no magnetic field term explicitly present in our model, $ $the
three states $\ket{T_{ij}^{\left(0,\pm\right)}}$ are degenerate in
energy {[}see Fig. \ref{fig:leveldiag}(c){]} and we may choose a
representative triplet state to calculate the singlet-triplet energy
splitting. Extensions to large parallel magnetic field cause no changes
for homogeneous $g$ factors throughout the impurity-dot system; inhomogeneous
corrections are considered at the end of this work.

We now calculate the energy shifts of the $\left(1,2,1\right)$ states
due to $H_{t}.$ For $S_{\textrm{tot}}=0,$ the matrix representation
of $H_{{\rm hub}}$ in the basis \{$\ket{\left(1,2,1\right);S_{LR},S_{11}},$
$\ket{\left(0,3,1\right);S_{R2},S_{11}},$ $\ket{\left(1,3,0\right);S_{L2},S_{11}},$
$\ket{\left(1,2^{\ast},1\right);S_{LR},S_{12}},$ $\ket{\left(1,2^{\ast},1\right);T_{LR},T_{12},+}$
\}, where $\ket{T_{LR},T_{12},+}\equiv\left(\ket{T_{LR}^{\left(0\right)},T_{12}^{\left(0\right)}}-\ket{T_{LR}^{\left(+\right)},T_{12}^{\left(-\right)}}-\ket{T_{LR}^{\left(-\right)},T_{12}^{\left(+\right)}}\right)/\sqrt{3}$
is symmetric with respect to exchange of the electron spin pairs $LR$
and $12,$ is given by 
\begin{equation}
H_{S}\equiv\left(\begin{array}{c|cc|cc}
0 & -t_{L2}^{\ast} & -t_{R2}^{\ast}\\
\hline -t_{L2} & \Delta_{L} &  & \frac{t_{L1}}{\sqrt{2}} & -\sqrt{\frac{3}{2}}t_{L1}\\
-t_{R2} &  & \Delta_{R} & \frac{t_{R1}}{\sqrt{2}} & \sqrt{\frac{3}{2}}t_{R1}\\
\hline  & \frac{t_{L1}^{\ast}}{\sqrt{2}} & \frac{t_{R1}^{\ast}}{\sqrt{2}} & \Delta_{J}\\
 & -\sqrt{\frac{3}{2}}t_{L1}^{\ast} & \sqrt{\frac{3}{2}}t_{R1}^{\ast} &  & \Delta_{M}
\end{array}\right).\label{eq:HS}
\end{equation}
For the $S_{\textrm{tot}}=1$ subspace in the basis \{$\ket{\left(1,2,1\right);T_{LR}^{\left(0\right)},S_{11}},$
$\ket{\left(0,3,1\right);T_{R2}^{\left(0\right)},S_{11}},$ $\ket{\left(1,3,0\right);T_{L2}^{\left(0\right)},S_{11}},$
$\ket{\left(1,2^{\ast},1\right);T_{LR}^{\left(0\right)},S_{12}},$
$\ket{\left(1,2^{\ast},1\right);S_{LR},T_{12}^{\left(0\right)}},$
$\ket{\left(1,2^{\ast},1\right);T_{LR},T_{12},-}$ \}, where $\ket{T_{LR},T_{12},-}\equiv\left(\ket{T_{LR}^{\left(+\right)},T_{12}^{\left(-\right)}}-\ket{T_{LR}^{\left(-\right)},T_{12}^{\left(+\right)}}\right)/\sqrt{2}$
is antisymmetric with respect to exchange of the electron spin pairs
$LR$ and $12,$ $H_{{\rm hub}}$ takes the form 
\begin{equation}
H_{T}\equiv\left(\begin{array}{c|cc|ccc}
0 & t_{L2}^{\ast} & -t_{R2}^{\ast}\\
\hline t_{L2} & \Delta_{L} &  & -\frac{t_{L1}}{\sqrt{2}} & \frac{t_{L1}}{\sqrt{2}} & t_{L1}\\
-t_{R2} &  & \Delta_{R} & \frac{t_{R1}}{\sqrt{2}} & \frac{t_{R1}}{\sqrt{2}} & -t_{R1}\\
\hline  & -\frac{t_{L1}^{\ast}}{\sqrt{2}} & \frac{t_{R1}^{\ast}}{\sqrt{2}} & \Delta_{J}\\
 & \frac{t_{L1}^{\ast}}{\sqrt{2}} & \frac{t_{R1}^{\ast}}{\sqrt{2}} &  & \Delta_{M}\\
 & t_{L1}^{\ast} & -t_{R1}^{\ast} &  &  & \Delta_{M}
\end{array}\right).\label{eq:HT}
\end{equation}
Using Eqs. \eqref{eq:HS} and \eqref{eq:HT}, we calculate the energy
shifts of $\ket{\left(1,2,1\right);S_{LR},S_{11}}$ and $\ket{\left(1,2,1\right);T_{LR}^{\left(0\right)},S_{11}}$
up to fourth order in $H_{t}$. We find that the first-order and third-order
corrections to the energy vanish, while the second-order shifts are
identical for both states$.$ The fourth-order shifts $\delta E_{S}^{\left(4\right)}$
and $\delta E_{T}^{\left(4\right)}$ are therefore the lowest-order
corrections that give rise to an energy splitting. The difference
$ $$\delta E_{T}^{\left(4\right)}-\delta E_{S}^{\left(4\right)}$
is the Heisenberg exchange coupling $J,$ which we find to be given
by 
\begin{equation}
J=-2\left(\frac{t_{R2}^{\ast}t_{R1}t_{L1}^{\ast}t_{L2}}{\Delta_{R}\Delta_{M}\Delta_{L}}+c.c.\right).\label{eq:J}
\end{equation}
This is the central result of our paper: using an initial singlet
configuration yields an RKKY-like interaction \cite{Braun2011}, including
both small-energy intermediate states ($\Delta_{M}$ being `small'
compared to the dot charging energy) and non-trivial interference
terms ($J$ depends on the phases of the tunneling terms in the presence
of the magnetic fields typically present in experiments).

Examining Eq.~\eqref{eq:J}, we first remark that $\Delta_{J},$
which differs from $\Delta_{M}$ by the intradot exchange splitting
$2J_{12},$ does not appear in this expression. From the dependence
of Eq. (\ref{eq:J}) on $\Delta_{L},$ $\Delta_{R},$ and $\Delta_{M}$,
we see that $J$ is inversely proportional to the energy detunings
$\epsilon_{2}-\epsilon_{L}$ and $\epsilon_{2}-\epsilon_{R}$ between
orbital 2 of the quantum dot and the impurity orbitals as well as
to the on-site energy difference $\epsilon_{2}-\epsilon_{1}$ between
the two levels of the quantum dot. As the detunings can be controlled
via the voltages applied to the dot and have a lower limit set only
by the tunnel coupling and magnetic field magnitudes, the strength
of the exchange coupling mediated by the two-level dot is highly tunable
and may be made large. This is in contrast to keeping a large detuning
to suppress sequential tunneling~\cite{Busl2013,Braakman2013}, which
limits the maximum achievable coupling strength.

We now turn to the phase dependence in Eq.~\eqref{eq:J}. The terms
correspond to two alternative pathways for the electrons which give
rise to the effective coupling $J$ (Fig. \ref{fig:vtp}); thus, the
interaction can have interference between these pathways, and their
non-trivial relative phase for finite magnetic fields leads to an
interaction strength that depends on the tunneling phase factors \cite{Braun2011}.
This provides a glimpse of the beginning of the expected sign fluctuations
in exchange for a true RKKY interaction, where the finite Fermi wave
vector $k_{F}$ of the two-electron Fermi `sea' matters. We note that
for phosphorus donor electrons in silicon, the tunneling amplitudes
also oscillate rapidly with the donor positions due to interference
between electronic states associated with different degenerate minima,
or valleys, existing in the conduction band \cite{Cullis1970,Koiller2001}.
This can be seen by taking $t_{ij}\propto\left\langle \psi_{i}\right.\ket{\psi_{j}}$
for $i=L,R$ and $j=1,2,$ where $\psi_{i,j}$ are superpositions
of orbital wave functions associated with each valley. The resulting
sinusoidal dependence of the tunneling amplitudes on the positions
of the donors relative to the dot center leads to a strong dependence
of the terms in Eq. \eqref{eq:J} on these relative positions. 

\emph{Charge noise and exchange gate fidelity}: Fluctuating electric
fields introduce variations in the parameters determining the effective
exchange $J$ in Eq. \eqref{eq:J} and consequently affect the operation
of exchange-based gates \cite{Loss1998,Hu2006,Taylor2007}. Here,
we consider the effects of classical charge noise on the detuning
parameters $\Delta_{\alpha}$ for $\alpha=L,M,R$ and calculate the
fidelity of the exchange gate $\hat{U}\left(\tau\right)=\exp\left(-iH_{{\rm exch}}\tau\right),$
where $H_{{\rm exch}}=-J\ket{S_{LR},S_{11}}\bra{S_{LR},S_{11}}$ and
$\ket{S_{LR},S_{11}}$ is the corrected state after elimination of
states outside the $\left(1,2,1\right)$ subspace (note that we suppress
the charge state in this notation, since the effective Hamiltonian
acts only in this subspace). Letting $\Delta_{\alpha}\rightarrow\Delta_{\alpha}+\delta_{\alpha},$
where $\delta_{\alpha}$ represents small fluctuations about the average
detuning $\Delta_{\alpha},$ and expanding to first order in $\delta_{\alpha}$
gives $J\rightarrow J'=J\left(1-\sum_{\alpha}\delta_{\alpha}/\Delta_{\alpha}\right)$.
We assume that the fluctuations $\delta_{\alpha}$ are independent
and described by Gaussian distributions $\text{\ensuremath{\rho_{\alpha}\left(\delta_{\alpha}\right)}}=e^{-\text{\ensuremath{\delta_{\alpha}^{2}}}/2\text{\ensuremath{\sigma_{\alpha}^{2}}}}/\sqrt{2\pi}\text{\ensuremath{\sigma_{\alpha}}}$
with charge noise standard deviations $\sigma_{\alpha}.$ The average
of the exchange gate over these fluctuations is then given by $\hat{U}'\left(\tau\right)={\bf 1}+\left(\langle e^{iJ'\tau}\rangle-1\right)\ket{S_{LR},S_{11}}\bra{S_{LR},S_{11}}$,
where $\left\langle e^{iJ'\tau}\right\rangle =e^{-\left(J^{2}\tau^{2}/2\right)\sum_{\alpha}\sigma_{\alpha}^{2}/\Delta_{\alpha}^{2}}e^{iJ\tau}.$
Note that the amplitude of $\left\langle e^{iJ'\tau}\right\rangle $
describes Gaussian decay of the form $e^{-\tau^{2}/T_{d}^{2}}$ with
a decay time $T_{d}=\left(1/J\right)\sqrt{2/\sum_{\alpha}\sigma_{\alpha}^{2}/\Delta_{\alpha}^{2}}$
\cite{Dial2013}.

\begin{figure}
\includegraphics[bb=20bp 20bp 280bp 280bp,width=2.75in]{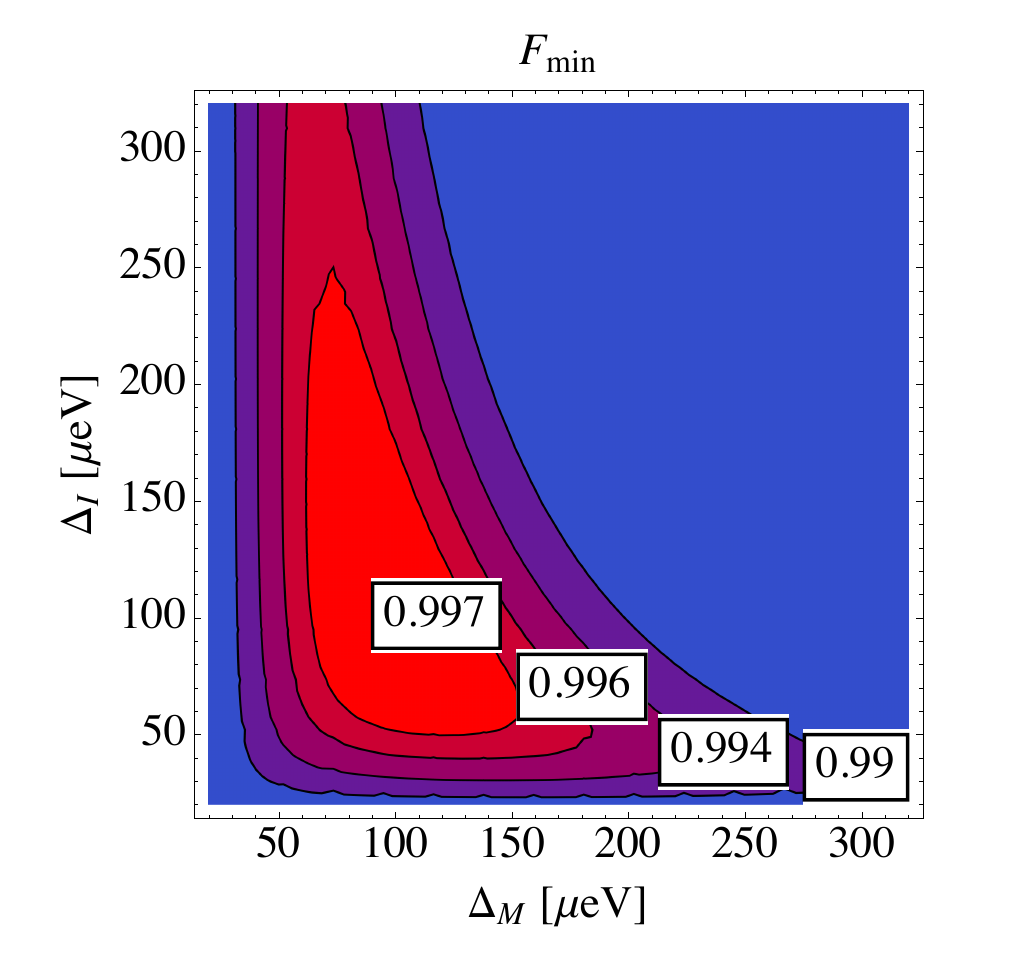}

\caption{\label{fig:Fmin} Minimum fidelity {[}Eq. \eqref{eq:Fmin}{]} of the
square-root-of-swap exchange gate $U_{{\rm sw}}^{1/2}\equiv\hat{U}\left(\pi/2J\right)$
as a function of the quantum dot level splitting $\Delta_{M}$ and
impurity-dot detunings $\Delta_{L}=\Delta_{R}\equiv\Delta_{I}$ for
$\sigma_{L}=\sigma_{R}=\sigma_{M}=2\ \mu{\rm eV},$ $T_{2}^{\ast}=1\ {\rm ms},$
and $t_{Li}=t_{Ri}=t=2\ \mu{\rm eV}.$ }
\end{figure}

We define the minimum gate fidelity as $F_{{\rm min}}\left(\tau\right)=e^{-\tau^{2}/T_{2}^{\ast2}}\left|\langle\psi_{\text{0}}|\hat{U}_{{\rm 0}}^{\dagger}\left(\tau\right)\hat{U}'\left(\tau\right)|\psi_{\text{0}}\rangle\right|^{2}$
\cite{Vandersypen2005}, where $\hat{U}_{0}\left(\tau\right)={\bf 1}+\left(e^{iJ\tau}-1\right)\ket{S_{LR},S_{11}}\bra{S_{LR},S_{11}}$
is the ideal gate and $\ket{\psi_{0}}=\left(\ket{T_{LR}^{\left(0\right)},S_{11}}+\ket{S_{LR},S_{11}}\right)/\sqrt{2}=\ket{\uparrow_{L}\downarrow_{R},S_{11}}$
is a state for which the exchange gate error is maximized. We also
include a factor $e^{-\tau^{2}/T_{2}^{\ast2}}$ to account for additional
decay characterized by a time $T_{2}^{\ast}$ over the gate duration
$\tau.$ Using the expression for $\left\langle e^{iJ'\tau}\right\rangle $,
we find 
\begin{equation}
F_{{\rm min}}\left(\tau\right)=\frac{e^{-\tau^{2}/T_{2}^{\ast2}}}{4}\left(1+e^{-\frac{1}{2}J^{2}\tau^{2}\sum_{\alpha}\frac{\sigma_{\alpha}^{2}}{\Delta_{\alpha}^{2}}}\right).\label{eq:Fmin}
\end{equation}
We plot this fidelity for the square-root-of-swap entangling gate
$U_{{\rm sw}}^{1/2}\equiv\hat{U}\left(\pi/2J\right)$ \cite{Loss1998}
as a function of the effective quantum dot level splitting $\Delta_{M}$
and symmetric effective impurity-dot detunings $\Delta_{L}=\Delta_{R}\equiv\Delta_{I}$
in Fig. \ref{fig:Fmin}, where we choose $\sigma_{L}=\sigma_{R}=\sigma_{M}=2\ \mu{\rm eV}$
and $T_{2}^{\ast}=1\ {\rm ms}$ \cite{Witzel2012}. For $\Delta_{M}=90\ \mu{\rm eV},$
$\Delta_{I}=60\ \mu{\rm eV,}$ and a tunnel coupling $t_{Li}=t_{Ri}=t=2\ \mu{\rm eV},$
which is relevant for phosphorus donors in silicon \cite{Morello2010,Pla2012},
we find $J=4t^{4}/\Delta_{R}\Delta_{M}\Delta_{L}=0.2\ {\rm neV}.$
This exchange coupling strength corresponds to a gate time $\tau_{{\rm gate}}=\pi/2J\approx5\ \mu{\rm s}$
and gate fidelity $F_{{\rm min}}\approx0.997.$ Thus, setting the
quantum dot level splitting and impurity-dot detunings to values within
an optimal range in principle enables high-fidelity exchange gates.

Finally, studies of exchange in multi-electron quantum dots \cite{Vorojtsov2004,Barnes2011,Higginbotham2014}
suggest that exchange coupling of the type discussed in the present
work, which is derived from tunneling via an excited orbital of a
multi-level quantum dot with lower-energy orbitals filled by electron
pairs, may exhibit increased robustness against fluctuations caused
by charge noise due to screening of the Coulomb interaction by the
paired ``core'' electrons already present in the dot. Varying the
number of electrons in the dot changes the spacing between the outermost
levels \cite{Hanson2007RMP} and consequently $\Delta_{M},$ so that
$J$ may be tuned in discrete steps. Provided this discrete level
description remains valid, the larger sizes associated with multi-electron
dots may also enable longer-range coupling.

\emph{Effects of inhomogeneous $g$ factors}: In the presence of an
external magnetic field, a difference in the $g$ factors of the impurities
and the quantum dot couples the $S_{{\rm tot}}=0$ and $S_{{\rm tot}}=1$
subspaces. To investigate the form of this coupling, we assume an
applied magnetic field ${\bf B}=B_{z}\hat{z}$ and add a magnetic
gradient term of the form 
\begin{equation}
H_{Z}=\frac{\Omega_{z}}{2}\sum_{i=1,2}\left(n_{i,\uparrow}-n_{i,\downarrow}\right)\label{eq:Hz}
\end{equation}
to the Hubbard Hamiltonian {[}Eqs. \eqref{eq:Hn} and \eqref{eq:Ht}{]},
where $\Omega_{z}\equiv\Delta g_{z}\mu_{B}B_{z}$ is the magnetic
field splitting due to a g-factor gradient $\Delta g_{z}$ parallel
to the external field {[}see Fig. \ref{fig:leveldiag}(c){]}. The
full Hamiltonian is then given by $H=H_{{\rm hub}}+H_{Z}=H_{n}+H_{t}+H_{Z}$
and acts in the combined space consisting of both the $S_{{\rm tot}}=0$
and $S_{{\rm tot}}=1$ subspaces. We transform to a basis which diagonalizes
$H_{0}\equiv H_{n}+H_{Z}$ and treat $H_{t}$ as a perturbation to
$H_{0}.$ Keeping terms up to second order in the tunneling amplitudes
and up to linear order in $\Omega_{z}$, we find that the correction
to the effective exchange Hamiltonian $H_{{\rm exch}}$ is given by
$H_{g}=f_{g}\left(\ket{T_{LR}^{\left(0\right)},S_{11}}\bra{S_{LR},S_{11}}+\ket{S_{LR},S_{11}}\bra{T_{LR}^{\left(0\right)},S_{11}}\right),$
where 
\begin{equation}
f_{g}=\frac{\Omega_{z}}{2}\left(\frac{\left|t_{L2}\right|^{2}}{\Delta_{L}^{2}}-\frac{\left|t_{R2}\right|^{2}}{\Delta_{R}^{2}}\right).\label{eq:fg}
\end{equation}
From this expression, we see that the effects of the $g$ factor inhomogeneity
described by Eq. \eqref{eq:Hz} can be eliminated up to first order
in $\Omega_{z}$ and second order in the tunneling amplitudes by choosing
$t_{L2},$ $t_{R2},$ $\Delta_{L}$ and $\Delta_{R}$ such that the
constraint $\Delta_{L}^{2}/\Delta_{R}^{2}=\left|t_{L2}\right|^{2}/\left|t_{R2}\right|^{2}$
is satisfied. Note that the preceding analysis assumes $\Omega_{z}<\Delta_{M,L,R},$
which sets an upper bound on $J$ {[}see Eq. \eqref{eq:J}{]}. For
impurity atoms with nonzero nuclear spin, hyperfine coupling represents
an additional source of magnetic gradients between the impurity and
dot electrons that may prove useful for alternative coupling schemes.
Indeed, for direct exchange coupling between two donor electron spins
in silicon, recent work \cite{Kalra2013arxiv} shows that a difference
in the hyperfine coupling between the donors enables two distinct
methods for realizing high-fidelity two-qubit gates. 

The validity of the toy model for the effective exchange coupling
considered in the present work is limited by the validity of the two-level
approximation for the mediator quantum dot in the presence of the
Coulomb interaction among the four electrons. Future work should consider
a detailed calculation of the effective exchange interaction mediated
by the two-level quantum dot in terms of the general form of the pairwise
Coulomb interaction and explore how this analysis may be extended
to gain insight into the form of the coupling mediated by a quantum
dot with more than two levels.

We thank A. Morello, W. M. Witzel, M. S. Carroll, L. I. Glazman, M.
D. Stewart, Jr., B. M. Anderson, F. R. Braakman, M. Friesen, and S.
N. Coppersmith for helpful discussions and valuable insights. This
work was supported by DARPA MTO and the NSF funded Physics Frontier
Center at the JQI. 

\bibliographystyle{apsrev}
\bibliography{TLQDC}

\begin{thebibliography}{54}
\expandafter\ifx\csname natexlab\endcsname\relax\def\natexlab#1{#1}\fi
\expandafter\ifx\csname bibnamefont\endcsname\relax
  \def\bibnamefont#1{#1}\fi
\expandafter\ifx\csname bibfnamefont\endcsname\relax
  \def\bibfnamefont#1{#1}\fi
\expandafter\ifx\csname citenamefont\endcsname\relax
  \def\citenamefont#1{#1}\fi
\expandafter\ifx\csname url\endcsname\relax
  \def\url#1{\texttt{#1}}\fi
\expandafter\ifx\csname urlprefix\endcsname\relax\def\urlprefix{URL }\fi
\providecommand{\bibinfo}[2]{#2}
\providecommand{\eprint}[2][]{\url{#2}}

\bibitem[{\citenamefont{Loss and DiVincenzo}(1998)}]{Loss1998}
\bibinfo{author}{\bibfnamefont{D.}~\bibnamefont{Loss}} \bibnamefont{and}
  \bibinfo{author}{\bibfnamefont{D.~P.} \bibnamefont{DiVincenzo}},
  \bibinfo{journal}{Phys. Rev. A} \textbf{\bibinfo{volume}{57}},
  \bibinfo{pages}{120} (\bibinfo{year}{1998}).

\bibitem[{\citenamefont{Kane}(1998)}]{Kane1998}
\bibinfo{author}{\bibfnamefont{B.~E.} \bibnamefont{Kane}},
  \bibinfo{journal}{Nature} \textbf{\bibinfo{volume}{393}},
  \bibinfo{pages}{133} (\bibinfo{year}{1998}).

\bibitem[{\citenamefont{Hanson et~al.}(2007)\citenamefont{Hanson, Kouwenhoven,
  Petta, Tarucha, and Vandersypen}}]{Hanson2007RMP}
\bibinfo{author}{\bibfnamefont{R.}~\bibnamefont{Hanson}},
  \bibinfo{author}{\bibfnamefont{L.~P.} \bibnamefont{Kouwenhoven}},
  \bibinfo{author}{\bibfnamefont{J.~R.} \bibnamefont{Petta}},
  \bibinfo{author}{\bibfnamefont{S.}~\bibnamefont{Tarucha}}, \bibnamefont{and}
  \bibinfo{author}{\bibfnamefont{L.~M.~K.} \bibnamefont{Vandersypen}},
  \bibinfo{journal}{Rev. Mod. Phys.} \textbf{\bibinfo{volume}{79}},
  \bibinfo{pages}{1217} (\bibinfo{year}{2007}).

\bibitem[{\citenamefont{Hanson and Awschalom}(2008)}]{Hanson2008}
\bibinfo{author}{\bibfnamefont{R.}~\bibnamefont{Hanson}} \bibnamefont{and}
  \bibinfo{author}{\bibfnamefont{D.~D.} \bibnamefont{Awschalom}},
  \bibinfo{journal}{Nature} \textbf{\bibinfo{volume}{453}},
  \bibinfo{pages}{1043} (\bibinfo{year}{2008}).

\bibitem[{\citenamefont{Awschalom et~al.}(2013)\citenamefont{Awschalom,
  Bassett, Dzurak, Hu, and Petta}}]{Awschalom2013}
\bibinfo{author}{\bibfnamefont{D.~D.} \bibnamefont{Awschalom}},
  \bibinfo{author}{\bibfnamefont{L.~C.} \bibnamefont{Bassett}},
  \bibinfo{author}{\bibfnamefont{A.~S.} \bibnamefont{Dzurak}},
  \bibinfo{author}{\bibfnamefont{E.~L.} \bibnamefont{Hu}}, \bibnamefont{and}
  \bibinfo{author}{\bibfnamefont{J.~R.} \bibnamefont{Petta}},
  \bibinfo{journal}{Science} \textbf{\bibinfo{volume}{339}},
  \bibinfo{pages}{1174} (\bibinfo{year}{2013}).

\bibitem[{\citenamefont{Zwanenburg et~al.}(2013)\citenamefont{Zwanenburg,
  Dzurak, Morello, Simmons, Hollenberg, Klimeck, Rogge, Coppersmith, and
  Eriksson}}]{Zwanenburg2013}
\bibinfo{author}{\bibfnamefont{F.~A.} \bibnamefont{Zwanenburg}},
  \bibinfo{author}{\bibfnamefont{A.~S.} \bibnamefont{Dzurak}},
  \bibinfo{author}{\bibfnamefont{A.}~\bibnamefont{Morello}},
  \bibinfo{author}{\bibfnamefont{M.~Y.} \bibnamefont{Simmons}},
  \bibinfo{author}{\bibfnamefont{L.~C.~L.} \bibnamefont{Hollenberg}},
  \bibinfo{author}{\bibfnamefont{G.}~\bibnamefont{Klimeck}},
  \bibinfo{author}{\bibfnamefont{S.}~\bibnamefont{Rogge}},
  \bibinfo{author}{\bibfnamefont{S.~N.} \bibnamefont{Coppersmith}},
  \bibnamefont{and} \bibinfo{author}{\bibfnamefont{M.~A.}
  \bibnamefont{Eriksson}}, \bibinfo{journal}{Rev. Mod. Phys.}
  \textbf{\bibinfo{volume}{85}}, \bibinfo{pages}{961} (\bibinfo{year}{2013}).

\bibitem[{\citenamefont{Vrijen et~al.}(2000)\citenamefont{Vrijen, Yablonovitch,
  Wang, Jiang, Balandin, Roychowdhury, Mor, and DiVincenzo}}]{Vrijen2000}
\bibinfo{author}{\bibfnamefont{R.}~\bibnamefont{Vrijen}},
  \bibinfo{author}{\bibfnamefont{E.}~\bibnamefont{Yablonovitch}},
  \bibinfo{author}{\bibfnamefont{K.}~\bibnamefont{Wang}},
  \bibinfo{author}{\bibfnamefont{H.~W.} \bibnamefont{Jiang}},
  \bibinfo{author}{\bibfnamefont{A.}~\bibnamefont{Balandin}},
  \bibinfo{author}{\bibfnamefont{V.}~\bibnamefont{Roychowdhury}},
  \bibinfo{author}{\bibfnamefont{T.}~\bibnamefont{Mor}}, \bibnamefont{and}
  \bibinfo{author}{\bibfnamefont{D.}~\bibnamefont{DiVincenzo}},
  \bibinfo{journal}{Phys. Rev. A} \textbf{\bibinfo{volume}{62}},
  \bibinfo{pages}{012306} (\bibinfo{year}{2000}).

\bibitem[{\citenamefont{Burkard et~al.}(1999)\citenamefont{Burkard, Loss, and
  DiVincenzo}}]{Burkard1999}
\bibinfo{author}{\bibfnamefont{G.}~\bibnamefont{Burkard}},
  \bibinfo{author}{\bibfnamefont{D.}~\bibnamefont{Loss}}, \bibnamefont{and}
  \bibinfo{author}{\bibfnamefont{D.~P.} \bibnamefont{DiVincenzo}},
  \bibinfo{journal}{Phys. Rev. B} \textbf{\bibinfo{volume}{59}},
  \bibinfo{pages}{2070} (\bibinfo{year}{1999}).

\bibitem[{\citenamefont{DiVincenzo et~al.}(2000)\citenamefont{DiVincenzo,
  Bacon, Kempe, Burkard, and Whaley}}]{DiVincenzo2000Nature}
\bibinfo{author}{\bibfnamefont{D.~P.} \bibnamefont{DiVincenzo}},
  \bibinfo{author}{\bibfnamefont{D.}~\bibnamefont{Bacon}},
  \bibinfo{author}{\bibfnamefont{J.}~\bibnamefont{Kempe}},
  \bibinfo{author}{\bibfnamefont{G.}~\bibnamefont{Burkard}}, \bibnamefont{and}
  \bibinfo{author}{\bibfnamefont{K.~B.} \bibnamefont{Whaley}},
  \bibinfo{journal}{Nature} \textbf{\bibinfo{volume}{408}},
  \bibinfo{pages}{339} (\bibinfo{year}{2000}).

\bibitem[{\citenamefont{Wu and Lidar}(2002)}]{Wu2002}
\bibinfo{author}{\bibfnamefont{L.-A.} \bibnamefont{Wu}} \bibnamefont{and}
  \bibinfo{author}{\bibfnamefont{D.~A.} \bibnamefont{Lidar}},
  \bibinfo{journal}{Phys. Rev. Lett.} \textbf{\bibinfo{volume}{88}},
  \bibinfo{pages}{207902} (\bibinfo{year}{2002}).

\bibitem[{\citenamefont{Taylor et~al.}(2005)\citenamefont{Taylor, Engel, Dur,
  Yacoby, Marcus, Zoller, and Lukin}}]{Taylor2005}
\bibinfo{author}{\bibfnamefont{J.~M.} \bibnamefont{Taylor}},
  \bibinfo{author}{\bibfnamefont{H.~A.} \bibnamefont{Engel}},
  \bibinfo{author}{\bibfnamefont{W.}~\bibnamefont{Dur}},
  \bibinfo{author}{\bibfnamefont{A.}~\bibnamefont{Yacoby}},
  \bibinfo{author}{\bibfnamefont{C.~M.} \bibnamefont{Marcus}},
  \bibinfo{author}{\bibfnamefont{P.}~\bibnamefont{Zoller}}, \bibnamefont{and}
  \bibinfo{author}{\bibfnamefont{M.~D.} \bibnamefont{Lukin}},
  \bibinfo{journal}{Nature Phys.} \textbf{\bibinfo{volume}{1}},
  \bibinfo{pages}{177} (\bibinfo{year}{2005}).

\bibitem[{\citenamefont{Doherty and Wardrop}(2013)}]{Doherty2013}
\bibinfo{author}{\bibfnamefont{A.~C.} \bibnamefont{Doherty}} \bibnamefont{and}
  \bibinfo{author}{\bibfnamefont{M.~P.} \bibnamefont{Wardrop}},
  \bibinfo{journal}{Phys. Rev. Lett.} \textbf{\bibinfo{volume}{111}},
  \bibinfo{pages}{050503} (\bibinfo{year}{2013}).

\bibitem[{\citenamefont{Taylor et~al.}(2013)\citenamefont{Taylor, Srinivasa,
  and Medford}}]{Taylor2013}
\bibinfo{author}{\bibfnamefont{J.~M.} \bibnamefont{Taylor}},
  \bibinfo{author}{\bibfnamefont{V.}~\bibnamefont{Srinivasa}},
  \bibnamefont{and} \bibinfo{author}{\bibfnamefont{J.}~\bibnamefont{Medford}},
  \bibinfo{journal}{Phys. Rev. Lett.} \textbf{\bibinfo{volume}{111}},
  \bibinfo{pages}{050502} (\bibinfo{year}{2013}).

\bibitem[{\citenamefont{Petta et~al.}(2005)\citenamefont{Petta, Johnson,
  Taylor, Laird, Yacoby, Lukin, Marcus, Hanson, and Gossard}}]{Petta2005}
\bibinfo{author}{\bibfnamefont{J.~R.} \bibnamefont{Petta}},
  \bibinfo{author}{\bibfnamefont{A.~C.} \bibnamefont{Johnson}},
  \bibinfo{author}{\bibfnamefont{J.~M.} \bibnamefont{Taylor}},
  \bibinfo{author}{\bibfnamefont{E.~A.} \bibnamefont{Laird}},
  \bibinfo{author}{\bibfnamefont{A.}~\bibnamefont{Yacoby}},
  \bibinfo{author}{\bibfnamefont{M.~D.} \bibnamefont{Lukin}},
  \bibinfo{author}{\bibfnamefont{C.~M.} \bibnamefont{Marcus}},
  \bibinfo{author}{\bibfnamefont{M.~P.} \bibnamefont{Hanson}},
  \bibnamefont{and} \bibinfo{author}{\bibfnamefont{A.~C.}
  \bibnamefont{Gossard}}, \bibinfo{journal}{Science}
  \textbf{\bibinfo{volume}{309}}, \bibinfo{pages}{2180} (\bibinfo{year}{2005}).

\bibitem[{\citenamefont{Maune et~al.}(2012)\citenamefont{Maune, Borselli,
  Huang, Ladd, Deelman, Holabird, Kiselev, Alvarado-Rodriguez, Ross, Schmitz
  et~al.}}]{Maune2012}
\bibinfo{author}{\bibfnamefont{B.~M.} \bibnamefont{Maune}},
  \bibinfo{author}{\bibfnamefont{M.~G.} \bibnamefont{Borselli}},
  \bibinfo{author}{\bibfnamefont{B.}~\bibnamefont{Huang}},
  \bibinfo{author}{\bibfnamefont{T.~D.} \bibnamefont{Ladd}},
  \bibinfo{author}{\bibfnamefont{P.~W.} \bibnamefont{Deelman}},
  \bibinfo{author}{\bibfnamefont{K.~S.} \bibnamefont{Holabird}},
  \bibinfo{author}{\bibfnamefont{A.~A.} \bibnamefont{Kiselev}},
  \bibinfo{author}{\bibfnamefont{I.}~\bibnamefont{Alvarado-Rodriguez}},
  \bibinfo{author}{\bibfnamefont{R.~S.} \bibnamefont{Ross}},
  \bibinfo{author}{\bibfnamefont{A.~E.} \bibnamefont{Schmitz}},
  \bibnamefont{et~al.}, \bibinfo{journal}{Nature}
  \textbf{\bibinfo{volume}{481}}, \bibinfo{pages}{344} (\bibinfo{year}{2012}).

\bibitem[{\citenamefont{Medford
  et~al.}(2013{\natexlab{a}})\citenamefont{Medford, Beil, Taylor, Bartlett,
  Doherty, Rashba, DiVincenzo, Lu, Gossard, and Marcus}}]{Medford2013NNano}
\bibinfo{author}{\bibfnamefont{J.}~\bibnamefont{Medford}},
  \bibinfo{author}{\bibfnamefont{J.}~\bibnamefont{Beil}},
  \bibinfo{author}{\bibfnamefont{J.~M.} \bibnamefont{Taylor}},
  \bibinfo{author}{\bibfnamefont{S.~D.} \bibnamefont{Bartlett}},
  \bibinfo{author}{\bibfnamefont{A.~C.} \bibnamefont{Doherty}},
  \bibinfo{author}{\bibfnamefont{E.~I.} \bibnamefont{Rashba}},
  \bibinfo{author}{\bibfnamefont{D.~P.} \bibnamefont{DiVincenzo}},
  \bibinfo{author}{\bibfnamefont{H.}~\bibnamefont{Lu}},
  \bibinfo{author}{\bibfnamefont{A.~C.} \bibnamefont{Gossard}},
  \bibnamefont{and} \bibinfo{author}{\bibfnamefont{C.~M.}
  \bibnamefont{Marcus}}, \bibinfo{journal}{Nature Nano.}
  \textbf{\bibinfo{volume}{8}}, \bibinfo{pages}{654}
  (\bibinfo{year}{2013}{\natexlab{a}}).

\bibitem[{\citenamefont{Medford
  et~al.}(2013{\natexlab{b}})\citenamefont{Medford, Beil, Taylor, Rashba, Lu,
  Gossard, and Marcus}}]{Medford2013}
\bibinfo{author}{\bibfnamefont{J.}~\bibnamefont{Medford}},
  \bibinfo{author}{\bibfnamefont{J.}~\bibnamefont{Beil}},
  \bibinfo{author}{\bibfnamefont{J.~M.} \bibnamefont{Taylor}},
  \bibinfo{author}{\bibfnamefont{E.~I.} \bibnamefont{Rashba}},
  \bibinfo{author}{\bibfnamefont{H.}~\bibnamefont{Lu}},
  \bibinfo{author}{\bibfnamefont{A.~C.} \bibnamefont{Gossard}},
  \bibnamefont{and} \bibinfo{author}{\bibfnamefont{C.~M.}
  \bibnamefont{Marcus}}, \bibinfo{journal}{Phys. Rev. Lett.}
  \textbf{\bibinfo{volume}{111}}, \bibinfo{pages}{050501}
  (\bibinfo{year}{2013}{\natexlab{b}}).

\bibitem[{\citenamefont{Herring and Flicker}(1964)}]{Herring1964}
\bibinfo{author}{\bibfnamefont{C.}~\bibnamefont{Herring}} \bibnamefont{and}
  \bibinfo{author}{\bibfnamefont{M.}~\bibnamefont{Flicker}},
  \bibinfo{journal}{Phys. Rev.} \textbf{\bibinfo{volume}{134}},
  \bibinfo{pages}{A362} (\bibinfo{year}{1964}).

\bibitem[{\citenamefont{Szkopek et~al.}(2006)\citenamefont{Szkopek, Boykin,
  Fan, Roychowdhury, Yablonovitch, Simms, Gyure, and Fong}}]{Szkopek2006}
\bibinfo{author}{\bibfnamefont{T.}~\bibnamefont{Szkopek}},
  \bibinfo{author}{\bibfnamefont{P.~O.} \bibnamefont{Boykin}},
  \bibinfo{author}{\bibfnamefont{H.}~\bibnamefont{Fan}},
  \bibinfo{author}{\bibfnamefont{V.~P.} \bibnamefont{Roychowdhury}},
  \bibinfo{author}{\bibfnamefont{E.}~\bibnamefont{Yablonovitch}},
  \bibinfo{author}{\bibfnamefont{G.}~\bibnamefont{Simms}},
  \bibinfo{author}{\bibfnamefont{M.}~\bibnamefont{Gyure}}, \bibnamefont{and}
  \bibinfo{author}{\bibfnamefont{B.}~\bibnamefont{Fong}},
  \bibinfo{journal}{IEEE Trans. Nanotechnol.} \textbf{\bibinfo{volume}{5}},
  \bibinfo{pages}{42} (\bibinfo{year}{2006}).

\bibitem[{\citenamefont{Stephens and Evans}(2009)}]{Stephens2009}
\bibinfo{author}{\bibfnamefont{A.~M.} \bibnamefont{Stephens}} \bibnamefont{and}
  \bibinfo{author}{\bibfnamefont{Z.~W.~E.} \bibnamefont{Evans}},
  \bibinfo{journal}{Phys. Rev. A} \textbf{\bibinfo{volume}{80}},
  \bibinfo{pages}{022313} (\bibinfo{year}{2009}).

\bibitem[{\citenamefont{Imamoglu et~al.}(1999)\citenamefont{Imamoglu,
  Awschalom, Burkard, DiVincenzo, Loss, Sherwin, and Small}}]{Imamoglu1999}
\bibinfo{author}{\bibfnamefont{A.}~\bibnamefont{Imamoglu}},
  \bibinfo{author}{\bibfnamefont{D.~D.} \bibnamefont{Awschalom}},
  \bibinfo{author}{\bibfnamefont{G.}~\bibnamefont{Burkard}},
  \bibinfo{author}{\bibfnamefont{D.~P.} \bibnamefont{DiVincenzo}},
  \bibinfo{author}{\bibfnamefont{D.}~\bibnamefont{Loss}},
  \bibinfo{author}{\bibfnamefont{M.}~\bibnamefont{Sherwin}}, \bibnamefont{and}
  \bibinfo{author}{\bibfnamefont{A.}~\bibnamefont{Small}},
  \bibinfo{journal}{Phys. Rev. Lett.} \textbf{\bibinfo{volume}{83}},
  \bibinfo{pages}{4204} (\bibinfo{year}{1999}).

\bibitem[{\citenamefont{Childress et~al.}(2004)\citenamefont{Childress,
  S\o{}rensen, and Lukin}}]{Childress2004}
\bibinfo{author}{\bibfnamefont{L.}~\bibnamefont{Childress}},
  \bibinfo{author}{\bibfnamefont{A.~S.} \bibnamefont{S\o{}rensen}},
  \bibnamefont{and} \bibinfo{author}{\bibfnamefont{M.~D.} \bibnamefont{Lukin}},
  \bibinfo{journal}{Phys. Rev. A} \textbf{\bibinfo{volume}{69}},
  \bibinfo{pages}{042302} (\bibinfo{year}{2004}).

\bibitem[{\citenamefont{Burkard and Imamoglu}(2006)}]{Burkard2006}
\bibinfo{author}{\bibfnamefont{G.}~\bibnamefont{Burkard}} \bibnamefont{and}
  \bibinfo{author}{\bibfnamefont{A.}~\bibnamefont{Imamoglu}},
  \bibinfo{journal}{Phys. Rev. B} \textbf{\bibinfo{volume}{74}},
  \bibinfo{pages}{041307} (\bibinfo{year}{2006}).

\bibitem[{\citenamefont{Taylor and Lukin}(2006)}]{Taylor2006}
\bibinfo{author}{\bibfnamefont{J.~M.} \bibnamefont{Taylor}} \bibnamefont{and}
  \bibinfo{author}{\bibfnamefont{M.~D.} \bibnamefont{Lukin}},
  \bibinfo{journal}{arXiv:cond-mat/0605144}  (\bibinfo{year}{2006}).

\bibitem[{\citenamefont{Frey et~al.}(2012)\citenamefont{Frey, Leek, Beck,
  Blais, Ihn, Ensslin, and Wallraff}}]{Frey2012}
\bibinfo{author}{\bibfnamefont{T.}~\bibnamefont{Frey}},
  \bibinfo{author}{\bibfnamefont{P.~J.} \bibnamefont{Leek}},
  \bibinfo{author}{\bibfnamefont{M.}~\bibnamefont{Beck}},
  \bibinfo{author}{\bibfnamefont{A.}~\bibnamefont{Blais}},
  \bibinfo{author}{\bibfnamefont{T.}~\bibnamefont{Ihn}},
  \bibinfo{author}{\bibfnamefont{K.}~\bibnamefont{Ensslin}}, \bibnamefont{and}
  \bibinfo{author}{\bibfnamefont{A.}~\bibnamefont{Wallraff}},
  \bibinfo{journal}{Phys. Rev. Lett.} \textbf{\bibinfo{volume}{108}},
  \bibinfo{pages}{046807} (\bibinfo{year}{2012}).

\bibitem[{\citenamefont{Petersson et~al.}(2012)\citenamefont{Petersson, McFaul,
  Schroer, Jung, Taylor, Houck, and Petta}}]{Petersson2012}
\bibinfo{author}{\bibfnamefont{K.~D.} \bibnamefont{Petersson}},
  \bibinfo{author}{\bibfnamefont{L.~W.} \bibnamefont{McFaul}},
  \bibinfo{author}{\bibfnamefont{M.~D.} \bibnamefont{Schroer}},
  \bibinfo{author}{\bibfnamefont{M.}~\bibnamefont{Jung}},
  \bibinfo{author}{\bibfnamefont{J.~M.} \bibnamefont{Taylor}},
  \bibinfo{author}{\bibfnamefont{A.~A.} \bibnamefont{Houck}}, \bibnamefont{and}
  \bibinfo{author}{\bibfnamefont{J.~R.} \bibnamefont{Petta}},
  \bibinfo{journal}{Nature} \textbf{\bibinfo{volume}{490}},
  \bibinfo{pages}{380} (\bibinfo{year}{2012}).

\bibitem[{\citenamefont{Trifunovic et~al.}(2012)\citenamefont{Trifunovic, Dial,
  Trif, Wootton, Abebe, Yacoby, and Loss}}]{Trifunovic2012}
\bibinfo{author}{\bibfnamefont{L.}~\bibnamefont{Trifunovic}},
  \bibinfo{author}{\bibfnamefont{O.}~\bibnamefont{Dial}},
  \bibinfo{author}{\bibfnamefont{M.}~\bibnamefont{Trif}},
  \bibinfo{author}{\bibfnamefont{J.~R.} \bibnamefont{Wootton}},
  \bibinfo{author}{\bibfnamefont{R.}~\bibnamefont{Abebe}},
  \bibinfo{author}{\bibfnamefont{A.}~\bibnamefont{Yacoby}}, \bibnamefont{and}
  \bibinfo{author}{\bibfnamefont{D.}~\bibnamefont{Loss}},
  \bibinfo{journal}{Phys. Rev. X} \textbf{\bibinfo{volume}{2}},
  \bibinfo{pages}{011006} (\bibinfo{year}{2012}).

\bibitem[{\citenamefont{Trifunovic et~al.}(2013)\citenamefont{Trifunovic,
  Pedrocchi, and Loss}}]{Trifunovic2013}
\bibinfo{author}{\bibfnamefont{L.}~\bibnamefont{Trifunovic}},
  \bibinfo{author}{\bibfnamefont{F.~L.} \bibnamefont{Pedrocchi}},
  \bibnamefont{and} \bibinfo{author}{\bibfnamefont{D.}~\bibnamefont{Loss}},
  \bibinfo{journal}{Phys. Rev. X} \textbf{\bibinfo{volume}{3}},
  \bibinfo{pages}{041023} (\bibinfo{year}{2013}).

\bibitem[{\citenamefont{Friesen et~al.}(2007)\citenamefont{Friesen, Biswas, Hu,
  and Lidar}}]{Friesen2007}
\bibinfo{author}{\bibfnamefont{M.}~\bibnamefont{Friesen}},
  \bibinfo{author}{\bibfnamefont{A.}~\bibnamefont{Biswas}},
  \bibinfo{author}{\bibfnamefont{X.}~\bibnamefont{Hu}}, \bibnamefont{and}
  \bibinfo{author}{\bibfnamefont{D.}~\bibnamefont{Lidar}},
  \bibinfo{journal}{Phys. Rev. Lett.} \textbf{\bibinfo{volume}{98}},
  \bibinfo{pages}{230503} (\bibinfo{year}{2007}).

\bibitem[{\citenamefont{Srinivasa et~al.}(2007)\citenamefont{Srinivasa, Levy,
  and Hellberg}}]{Srinivasa2007}
\bibinfo{author}{\bibfnamefont{V.}~\bibnamefont{Srinivasa}},
  \bibinfo{author}{\bibfnamefont{J.}~\bibnamefont{Levy}}, \bibnamefont{and}
  \bibinfo{author}{\bibfnamefont{C.~S.} \bibnamefont{Hellberg}},
  \bibinfo{journal}{Phys. Rev. B} \textbf{\bibinfo{volume}{76}},
  \bibinfo{pages}{094411} (\bibinfo{year}{2007}).

\bibitem[{\citenamefont{Oh et~al.}(2010)\citenamefont{Oh, Friesen, and
  Hu}}]{Oh2010}
\bibinfo{author}{\bibfnamefont{S.}~\bibnamefont{Oh}},
  \bibinfo{author}{\bibfnamefont{M.}~\bibnamefont{Friesen}}, \bibnamefont{and}
  \bibinfo{author}{\bibfnamefont{X.}~\bibnamefont{Hu}}, \bibinfo{journal}{Phys.
  Rev. B} \textbf{\bibinfo{volume}{82}}, \bibinfo{pages}{140403}
  (\bibinfo{year}{2010}).

\bibitem[{\citenamefont{Marcos et~al.}(2010)\citenamefont{Marcos, Wubs, Taylor,
  Aguado, Lukin, and S\o{}rensen}}]{Marcos2010}
\bibinfo{author}{\bibfnamefont{D.}~\bibnamefont{Marcos}},
  \bibinfo{author}{\bibfnamefont{M.}~\bibnamefont{Wubs}},
  \bibinfo{author}{\bibfnamefont{J.~M.} \bibnamefont{Taylor}},
  \bibinfo{author}{\bibfnamefont{R.}~\bibnamefont{Aguado}},
  \bibinfo{author}{\bibfnamefont{M.~D.} \bibnamefont{Lukin}}, \bibnamefont{and}
  \bibinfo{author}{\bibfnamefont{A.~S.} \bibnamefont{S\o{}rensen}},
  \bibinfo{journal}{Phys. Rev. Lett.} \textbf{\bibinfo{volume}{105}},
  \bibinfo{pages}{210501} (\bibinfo{year}{2010}).

\bibitem[{\citenamefont{Leijnse and Flensberg}(2013)}]{Leijnse2013}
\bibinfo{author}{\bibfnamefont{M.}~\bibnamefont{Leijnse}} \bibnamefont{and}
  \bibinfo{author}{\bibfnamefont{K.}~\bibnamefont{Flensberg}},
  \bibinfo{journal}{Phys. Rev. Lett.} \textbf{\bibinfo{volume}{111}},
  \bibinfo{pages}{060501} (\bibinfo{year}{2013}).

\bibitem[{\citenamefont{Lehmann et~al.}(2007)\citenamefont{Lehmann,
  Gaita-Arino, Coronado, and Loss}}]{Lehmann2007}
\bibinfo{author}{\bibfnamefont{J.}~\bibnamefont{Lehmann}},
  \bibinfo{author}{\bibfnamefont{A.}~\bibnamefont{Gaita-Arino}},
  \bibinfo{author}{\bibfnamefont{E.}~\bibnamefont{Coronado}}, \bibnamefont{and}
  \bibinfo{author}{\bibfnamefont{D.}~\bibnamefont{Loss}},
  \bibinfo{journal}{Nature Nano.} \textbf{\bibinfo{volume}{2}},
  \bibinfo{pages}{312} (\bibinfo{year}{2007}).

\bibitem[{\citenamefont{Busl et~al.}(2013)\citenamefont{Busl, Granger,
  Gaudreau, Sanchez, Kam, Pioro-Ladriere, Studenikin, Zawadzki, Wasilewski,
  Sachrajda et~al.}}]{Busl2013}
\bibinfo{author}{\bibfnamefont{M.}~\bibnamefont{Busl}},
  \bibinfo{author}{\bibfnamefont{G.}~\bibnamefont{Granger}},
  \bibinfo{author}{\bibfnamefont{L.}~\bibnamefont{Gaudreau}},
  \bibinfo{author}{\bibfnamefont{R.}~\bibnamefont{Sanchez}},
  \bibinfo{author}{\bibfnamefont{A.}~\bibnamefont{Kam}},
  \bibinfo{author}{\bibfnamefont{M.}~\bibnamefont{Pioro-Ladriere}},
  \bibinfo{author}{\bibfnamefont{S.~A.} \bibnamefont{Studenikin}},
  \bibinfo{author}{\bibfnamefont{P.}~\bibnamefont{Zawadzki}},
  \bibinfo{author}{\bibfnamefont{Z.~R.} \bibnamefont{Wasilewski}},
  \bibinfo{author}{\bibfnamefont{A.~S.} \bibnamefont{Sachrajda}},
  \bibnamefont{et~al.}, \bibinfo{journal}{Nature Nano.}
  \textbf{\bibinfo{volume}{8}}, \bibinfo{pages}{261} (\bibinfo{year}{2013}).

\bibitem[{\citenamefont{Braakman et~al.}(2013)\citenamefont{Braakman,
  Barthelemy, Reichl, Wegscheider, and Vandersypen}}]{Braakman2013}
\bibinfo{author}{\bibfnamefont{F.~R.} \bibnamefont{Braakman}},
  \bibinfo{author}{\bibfnamefont{P.}~\bibnamefont{Barthelemy}},
  \bibinfo{author}{\bibfnamefont{C.}~\bibnamefont{Reichl}},
  \bibinfo{author}{\bibfnamefont{W.}~\bibnamefont{Wegscheider}},
  \bibnamefont{and} \bibinfo{author}{\bibfnamefont{L.~M.~K.}
  \bibnamefont{Vandersypen}}, \bibinfo{journal}{Nature Nano.}
  \textbf{\bibinfo{volume}{8}}, \bibinfo{pages}{432} (\bibinfo{year}{2013}).

\bibitem[{\citenamefont{Recher et~al.}(2001)\citenamefont{Recher, Loss, and
  Levy}}]{Recher2001}
\bibinfo{author}{\bibfnamefont{P.}~\bibnamefont{Recher}},
  \bibinfo{author}{\bibfnamefont{D.}~\bibnamefont{Loss}}, \bibnamefont{and}
  \bibinfo{author}{\bibfnamefont{J.}~\bibnamefont{Levy}},
  \emph{\bibinfo{title}{Macroscopic Quantum Coherence and Quantum Computing}}
  (\bibinfo{publisher}{Kluwer Academic}, \bibinfo{year}{2001}), pp.
  \bibinfo{pages}{293--306}.

\bibitem[{\citenamefont{Craig et~al.}(2004)\citenamefont{Craig, Taylor, Lester,
  Marcus, Hanson, and Gossard}}]{Craig2004}
\bibinfo{author}{\bibfnamefont{N.~J.} \bibnamefont{Craig}},
  \bibinfo{author}{\bibfnamefont{J.~M.} \bibnamefont{Taylor}},
  \bibinfo{author}{\bibfnamefont{E.~A.} \bibnamefont{Lester}},
  \bibinfo{author}{\bibfnamefont{C.~M.} \bibnamefont{Marcus}},
  \bibinfo{author}{\bibfnamefont{M.~P.} \bibnamefont{Hanson}},
  \bibnamefont{and} \bibinfo{author}{\bibfnamefont{A.~C.}
  \bibnamefont{Gossard}}, \bibinfo{journal}{Science}
  \textbf{\bibinfo{volume}{304}}, \bibinfo{pages}{565} (\bibinfo{year}{2004}).

\bibitem[{\citenamefont{Morello et~al.}(2010)\citenamefont{Morello, Pla,
  Zwanenburg, Chan, Tan, Huebl, Mottonen, Nugroho, Yang, van Donkelaar
  et~al.}}]{Morello2010}
\bibinfo{author}{\bibfnamefont{A.}~\bibnamefont{Morello}},
  \bibinfo{author}{\bibfnamefont{J.~J.} \bibnamefont{Pla}},
  \bibinfo{author}{\bibfnamefont{F.~A.} \bibnamefont{Zwanenburg}},
  \bibinfo{author}{\bibfnamefont{K.~W.} \bibnamefont{Chan}},
  \bibinfo{author}{\bibfnamefont{K.~Y.} \bibnamefont{Tan}},
  \bibinfo{author}{\bibfnamefont{H.}~\bibnamefont{Huebl}},
  \bibinfo{author}{\bibfnamefont{M.}~\bibnamefont{Mottonen}},
  \bibinfo{author}{\bibfnamefont{C.~D.} \bibnamefont{Nugroho}},
  \bibinfo{author}{\bibfnamefont{C.}~\bibnamefont{Yang}},
  \bibinfo{author}{\bibfnamefont{J.~A.} \bibnamefont{van Donkelaar}},
  \bibnamefont{et~al.}, \bibinfo{journal}{Nature}
  \textbf{\bibinfo{volume}{467}}, \bibinfo{pages}{687} (\bibinfo{year}{2010}).

\bibitem[{\citenamefont{Pla et~al.}(2012)\citenamefont{Pla, Tan, Dehollain,
  Lim, Morton, Jamieson, Dzurak, and Morello}}]{Pla2012}
\bibinfo{author}{\bibfnamefont{J.~J.} \bibnamefont{Pla}},
  \bibinfo{author}{\bibfnamefont{K.~Y.} \bibnamefont{Tan}},
  \bibinfo{author}{\bibfnamefont{J.~P.} \bibnamefont{Dehollain}},
  \bibinfo{author}{\bibfnamefont{W.~H.} \bibnamefont{Lim}},
  \bibinfo{author}{\bibfnamefont{J.~J.~L.} \bibnamefont{Morton}},
  \bibinfo{author}{\bibfnamefont{D.~N.} \bibnamefont{Jamieson}},
  \bibinfo{author}{\bibfnamefont{A.~S.} \bibnamefont{Dzurak}},
  \bibnamefont{and} \bibinfo{author}{\bibfnamefont{A.}~\bibnamefont{Morello}},
  \bibinfo{journal}{Nature} \textbf{\bibinfo{volume}{489}},
  \bibinfo{pages}{541} (\bibinfo{year}{2012}).

\bibitem[{\citenamefont{Korkusinski et~al.}(2007)\citenamefont{Korkusinski,
  Gimenez, Hawrylak, Gaudreau, Studenikin, and Sachrajda}}]{Korkusinski2007}
\bibinfo{author}{\bibfnamefont{M.}~\bibnamefont{Korkusinski}},
  \bibinfo{author}{\bibfnamefont{I.~P.} \bibnamefont{Gimenez}},
  \bibinfo{author}{\bibfnamefont{P.}~\bibnamefont{Hawrylak}},
  \bibinfo{author}{\bibfnamefont{L.}~\bibnamefont{Gaudreau}},
  \bibinfo{author}{\bibfnamefont{S.~A.} \bibnamefont{Studenikin}},
  \bibnamefont{and} \bibinfo{author}{\bibfnamefont{A.~S.}
  \bibnamefont{Sachrajda}}, \bibinfo{journal}{Phys. Rev. B}
  \textbf{\bibinfo{volume}{75}}, \bibinfo{pages}{115301}
  (\bibinfo{year}{2007}).

\bibitem[{\citenamefont{Hsieh et~al.}(2012)\citenamefont{Hsieh, Shim, and
  Hawrylak}}]{Hsieh2012}
\bibinfo{author}{\bibfnamefont{C.-Y.} \bibnamefont{Hsieh}},
  \bibinfo{author}{\bibfnamefont{Y.-P.} \bibnamefont{Shim}}, \bibnamefont{and}
  \bibinfo{author}{\bibfnamefont{P.}~\bibnamefont{Hawrylak}},
  \bibinfo{journal}{Phys. Rev. B} \textbf{\bibinfo{volume}{85}},
  \bibinfo{pages}{085309} (\bibinfo{year}{2012}).

\bibitem[{\citenamefont{Braun et~al.}(2011)\citenamefont{Braun, Struck, and
  Burkard}}]{Braun2011}
\bibinfo{author}{\bibfnamefont{M.}~\bibnamefont{Braun}},
  \bibinfo{author}{\bibfnamefont{P.~R.} \bibnamefont{Struck}},
  \bibnamefont{and} \bibinfo{author}{\bibfnamefont{G.}~\bibnamefont{Burkard}},
  \bibinfo{journal}{Phys. Rev. B} \textbf{\bibinfo{volume}{84}},
  \bibinfo{pages}{115445} (\bibinfo{year}{2011}).

\bibitem[{\citenamefont{Cullis and Marko}(1970)}]{Cullis1970}
\bibinfo{author}{\bibfnamefont{P.~R.} \bibnamefont{Cullis}} \bibnamefont{and}
  \bibinfo{author}{\bibfnamefont{J.~R.} \bibnamefont{Marko}},
  \bibinfo{journal}{Phys. Rev. B} \textbf{\bibinfo{volume}{1}},
  \bibinfo{pages}{632} (\bibinfo{year}{1970}).

\bibitem[{\citenamefont{Koiller et~al.}(2001)\citenamefont{Koiller, Hu, and
  Das~Sarma}}]{Koiller2001}
\bibinfo{author}{\bibfnamefont{B.}~\bibnamefont{Koiller}},
  \bibinfo{author}{\bibfnamefont{X.}~\bibnamefont{Hu}}, \bibnamefont{and}
  \bibinfo{author}{\bibfnamefont{S.}~\bibnamefont{Das~Sarma}},
  \bibinfo{journal}{Phys. Rev. Lett.} \textbf{\bibinfo{volume}{88}},
  \bibinfo{pages}{027903} (\bibinfo{year}{2001}).

\bibitem[{\citenamefont{Hu and Das~Sarma}(2006)}]{Hu2006}
\bibinfo{author}{\bibfnamefont{X.}~\bibnamefont{Hu}} \bibnamefont{and}
  \bibinfo{author}{\bibfnamefont{S.}~\bibnamefont{Das~Sarma}},
  \bibinfo{journal}{Phys. Rev. Lett.} \textbf{\bibinfo{volume}{96}},
  \bibinfo{pages}{100501} (\bibinfo{year}{2006}).

\bibitem[{\citenamefont{Taylor et~al.}(2007)\citenamefont{Taylor, Petta,
  Johnson, Yacoby, Marcus, and Lukin}}]{Taylor2007}
\bibinfo{author}{\bibfnamefont{J.~M.} \bibnamefont{Taylor}},
  \bibinfo{author}{\bibfnamefont{J.~R.} \bibnamefont{Petta}},
  \bibinfo{author}{\bibfnamefont{A.~C.} \bibnamefont{Johnson}},
  \bibinfo{author}{\bibfnamefont{A.}~\bibnamefont{Yacoby}},
  \bibinfo{author}{\bibfnamefont{C.~M.} \bibnamefont{Marcus}},
  \bibnamefont{and} \bibinfo{author}{\bibfnamefont{M.~D.} \bibnamefont{Lukin}},
  \bibinfo{journal}{Phys. Rev. B} \textbf{\bibinfo{volume}{76}},
  \bibinfo{pages}{035315} (\bibinfo{year}{2007}).

\bibitem[{\citenamefont{Dial et~al.}(2013)\citenamefont{Dial, Shulman, Harvey,
  Bluhm, Umansky, and Yacoby}}]{Dial2013}
\bibinfo{author}{\bibfnamefont{O.~E.} \bibnamefont{Dial}},
  \bibinfo{author}{\bibfnamefont{M.~D.} \bibnamefont{Shulman}},
  \bibinfo{author}{\bibfnamefont{S.~P.} \bibnamefont{Harvey}},
  \bibinfo{author}{\bibfnamefont{H.}~\bibnamefont{Bluhm}},
  \bibinfo{author}{\bibfnamefont{V.}~\bibnamefont{Umansky}}, \bibnamefont{and}
  \bibinfo{author}{\bibfnamefont{A.}~\bibnamefont{Yacoby}},
  \bibinfo{journal}{Phys. Rev. Lett.} \textbf{\bibinfo{volume}{110}},
  \bibinfo{pages}{146804} (\bibinfo{year}{2013}).

\bibitem[{\citenamefont{Vandersypen and Chuang}(2005)}]{Vandersypen2005}
\bibinfo{author}{\bibfnamefont{L.~M.~K.} \bibnamefont{Vandersypen}}
  \bibnamefont{and} \bibinfo{author}{\bibfnamefont{I.~L.}
  \bibnamefont{Chuang}}, \bibinfo{journal}{Rev. Mod. Phys.}
  \textbf{\bibinfo{volume}{76}}, \bibinfo{pages}{1037} (\bibinfo{year}{2005}).

\bibitem[{\citenamefont{Witzel et~al.}(2012)\citenamefont{Witzel, Carroll,
  Cywi\ifmmode~\acute{n}\else \'{n}\fi{}ski, and Das~Sarma}}]{Witzel2012}
\bibinfo{author}{\bibfnamefont{W.~M.} \bibnamefont{Witzel}},
  \bibinfo{author}{\bibfnamefont{M.~S.} \bibnamefont{Carroll}},
  \bibinfo{author}{\bibfnamefont{L.}~\bibnamefont{Cywi\ifmmode~\acute{n}\else
  \'{n}\fi{}ski}}, \bibnamefont{and}
  \bibinfo{author}{\bibfnamefont{S.}~\bibnamefont{Das~Sarma}},
  \bibinfo{journal}{Phys. Rev. B} \textbf{\bibinfo{volume}{86}},
  \bibinfo{pages}{035452} (\bibinfo{year}{2012}).

\bibitem[{\citenamefont{Vorojtsov et~al.}(2004)\citenamefont{Vorojtsov,
  Mucciolo, and Baranger}}]{Vorojtsov2004}
\bibinfo{author}{\bibfnamefont{S.}~\bibnamefont{Vorojtsov}},
  \bibinfo{author}{\bibfnamefont{E.~R.} \bibnamefont{Mucciolo}},
  \bibnamefont{and} \bibinfo{author}{\bibfnamefont{H.~U.}
  \bibnamefont{Baranger}}, \bibinfo{journal}{Phys. Rev. B}
  \textbf{\bibinfo{volume}{69}}, \bibinfo{pages}{115329}
  (\bibinfo{year}{2004}).

\bibitem[{\citenamefont{Barnes et~al.}(2011)\citenamefont{Barnes, Kestner,
  Nguyen, and Das~Sarma}}]{Barnes2011}
\bibinfo{author}{\bibfnamefont{E.}~\bibnamefont{Barnes}},
  \bibinfo{author}{\bibfnamefont{J.~P.} \bibnamefont{Kestner}},
  \bibinfo{author}{\bibfnamefont{N.~T.~T.} \bibnamefont{Nguyen}},
  \bibnamefont{and}
  \bibinfo{author}{\bibfnamefont{S.}~\bibnamefont{Das~Sarma}},
  \bibinfo{journal}{Phys. Rev. B} \textbf{\bibinfo{volume}{84}},
  \bibinfo{pages}{235309} (\bibinfo{year}{2011}).

\bibitem[{\citenamefont{Higginbotham et~al.}(2014)\citenamefont{Higginbotham,
  Kuemmeth, Hanson, Gossard, and Marcus}}]{Higginbotham2014}
\bibinfo{author}{\bibfnamefont{A.~P.} \bibnamefont{Higginbotham}},
  \bibinfo{author}{\bibfnamefont{F.}~\bibnamefont{Kuemmeth}},
  \bibinfo{author}{\bibfnamefont{M.~P.} \bibnamefont{Hanson}},
  \bibinfo{author}{\bibfnamefont{A.~C.} \bibnamefont{Gossard}},
  \bibnamefont{and} \bibinfo{author}{\bibfnamefont{C.~M.}
  \bibnamefont{Marcus}}, \bibinfo{journal}{Phys. Rev. Lett.}
  \textbf{\bibinfo{volume}{112}}, \bibinfo{pages}{026801}
  (\bibinfo{year}{2014}).

\bibitem[{\citenamefont{Kalra et~al.}(2013)\citenamefont{Kalra, Laucht, Hill,
  and Morello}}]{Kalra2013arxiv}
\bibinfo{author}{\bibfnamefont{R.}~\bibnamefont{Kalra}},
  \bibinfo{author}{\bibfnamefont{A.}~\bibnamefont{Laucht}},
  \bibinfo{author}{\bibfnamefont{C.}~\bibnamefont{Hill}}, \bibnamefont{and}
  \bibinfo{author}{\bibfnamefont{A.}~\bibnamefont{Morello}},
  \bibinfo{journal}{arXiv:1312.2197}  (\bibinfo{year}{2013}).

\end{thebibliography}

\end{document}